# Accuracy and Resiliency of Analog Compute-in-Memory Inference Engines

Zhe Wan
> Electrical and Computer Engineering, University of California, Los Angeles, Los Angeles, CA, USA, z.wan@ucla.edu

Tianyi Wang
> Electrical and Computer Engineering, University of California, Los Angeles, Los Angeles, CA, USA, tianyiw@ucla.edu

Yiming Zhou
> Electrical and Computer Engineering, University of California, Los Angeles, Los Angeles, CA, USA, yimingz0416@ucla.edu

Subramanian S. Iyer
> Electrical and Computer Engineering, University of California, Los Angeles, Los Angeles, CA, USA, s.s.iyer@ucla.edu

Vwani P. Roychowdhury
> Electrical and Computer Engineering, University of California, Los Angeles, Los Angeles, CA, USA, vwani@ucla.edu

**ABSTRACT**

Recently, analog compute-in-memory (CIM) architectures based on emerging analog non-volatile memory (NVM) technologies have been explored for deep neural networks (DNN) to improve energy efficiency. Such architectures, however, leverage charge conservation, an operation with infinite resolution, and thus are susceptible to errors. The computations in DNN realized by analog NVM thus have high uncertainty due to the device stochasticity. Several reports have demonstrated the use of analog NVM for CIM in a limited scale. It is unclear whether the uncertainties in computations will prohibit large-scale DNNs. To explore this critical issue of scalability, this paper first presents a simulation framework to evaluate the feasibility of large-scale DNNs based on CIM architecture and analog NVM. Simulation results show that *DNNs trained for high-precision digital computing engines are not resilient* against the uncertainty of the analog NVM devices. To avoid such catastrophic failures, this paper introduces the analog floating-point representation for the DNN, and the Hessian-Aware Stochastic Gradient Descent (HA-SGD) training algorithm to enhance the inference accuracy of trained DNNs. As a result of such enhancements, *DNNs such as Wide ResNets for CIFAR-100 image recognition problem are demonstrated to have significant performance improvements in accuracy without adding cost to the inference hardware*.

# 1  Introduction

Deep learning based on deep neural networks (DNNs) has shown promising results in numerous applications such as computer vision, speech recognition and natural language processing [13]. While DNNs derive their inspiration from the brain, which is an analog system, they are primarily executed on digital machines, where information and computation are both digitized (e.g., with 32-bit floating-point precision) and usually operated within a von Neumann architecture. Although the von Neumann architecture has been prevalent and successful under the rapid development of the Moore's Law, it has encountered the von Neumann bottleneck -- a limitation of the bandwidth between the processors and the memory [1]. This von Neumann bottleneck is perhaps most evident in the execution of DNNs: They require a heavy workload of vector-matrix multiplication (VMM), whose operands can be very large in size. The frequent fetching of the operands due to VMMs makes neural networks computation data-intensive [20], and therefore particularly susceptible to the von Neumann bottleneck.

Many application-specific integrated-circuit (ASIC) designs based on CMOS technology have been developed to improve system performance by spreading out processing elements to enable parallel processing, increasing on-chip memory and optimizing dataflow to maximize throughput [6, 11]. Some other implementations compress the data by leveraging the sparsity of the synaptic weights of the neural networks. These solutions alleviate the von Neumann bottleneck, but the memory bandwidth is still a limiting factor.

To further address the von Neumann bottleneck, computing-in-memory (CIM) architectures are proposed for both digital and analog memories [12, 14, 24]. Digital CIM architectures are based on SRAM with a modified sense amplifier to read the current as a result of summation by Kirchhoff current law. Analog CIM architectures use analog non-volatile memories (ANVM) as synaptic weights, and have emerged as one of the main candidate technologies with the potential to improve energy-efficiency and throughput for DNN operations by several orders of magnitude. Preliminary results have been demonstrated on various analog devices. Emerging non-volatile memories (NVM), including charge-trap transistor (CTT) [7], Flash [8], phase-change memory (PCM) [21], resistive random-access memory (RRAM) [3], and spin-toque transfer magnetic RAM (STT-MRAM) [18] become promising candidates for mixed-signal CIM architecture.

One important aspect of the ANVM-based DNN accelerators is the imprecision of the ANVM devices, i.e., the stochastic variations of device resistance. The computation inside the ANVM arrays leverages natural laws (i.e., Kirchoffs Law, Ohm's Law) instead of numerical arithmetic laws for the traditional digital systems. Deviation of ANVM device resistance due to programming error and device instability directly leads to errors in the computations and could degrade the accuracy of the DNN significantly.

In this paper, we first present a simulation framework to estimate the effect of such analog uncertainties on the performance of DNNs. Our simulations show that large-scale DNNs that are conventionally trained for computing on Digital machine show catastrophic degradations in performance, when computed on ANVM-based DNN accelerators. To alleviate such scalability issues, we propose simultaneous enhancements in both the CIM architecture and in the training algorithms. For example, *at the training level*, we propose a hardware-aware training methodology to enhance the resiliency of the neural network against analog errors in a full analog CIM architecture. We use an architecture where each synaptic weight is implemented by one analog cell where we can model the network error due to device uncertainty with a continuous distribution. To minimize the degradation of the network due to this error, we optimize the Hessian of its cost function (i.e., drive the optimal weights to a point with not only a low loss-function value, but also around which the loss landscape is flat, or the Hessian's norm is small), by using the proposed Hessian-aware Stochastic Gradient Descent (HASGD) training algorithm. As a result, the trained model becomes more resilient to the device uncertainty without extra cost of the inference hardware. At the architectural level, we introduce the analog floating-point scheme to maximize the usable memory window of the ANVM cells to make the neural network more resilient to device errors. Furthermore, we explore the effect of digitization in the ANVM engines and quantify the network resiliency improvement as savings from the digitization effort. This is critical for ANVM-based DNN engines since the high area and energy cost of the analog-to-digital converters (ADCs) can undermine the advantages of such engines significantly.

The proposed method comes in naturally due to the proposed hardware architecture and does not require extra overhead in the hardware design or operation. We observe that the hardware-aware methodology increase the neural network resiliency by up to about 40% increase of network accuracy depending on the application, and more than 50% if low-resolution ADCs are used in the CIM architecture.

## 2  Related Work

While some ANVM-based neural network inference engines have been demonstrated [3, 8], they are still limited in scale, focusing on small size neural networks (e.g., multi-layer perceptron) and simple problems (e.g., digit recognition). While the hardware can be readily scaled if the device fabrication process can be integrated with standard CMOS technologies, it is not sure whether the neural networks are resilient to the intrinsically uncertain ANVM devices. Feasibility studies of such engines after scaling are mostly done in simulations. Simulators are developed to evaluate the performance of the CIM engine based on ANVM for DNN applications. For example, Chen et al. presented the NeuroSim simulator to evaluate the effect of errors from the analog synaptic weights and the circuits [4]. Two major sources of errors are investigated for inference engines: (1) the error of the analog devices, which are used as the synaptic weights of the CIM engine, and (2) the error of the peripheral circuits (e.g., charge integrator, analog-to-digital converter, etc.).

Hardware-based techniques are proposed to improve the robustness of DNN deployed in analog/mixed-signal CIM engine. Lin et al. proposed another simulation framework to model the impact of noise on the accuracy of RRAM based DNN accelerator, and a workload-dependent sensing scheme is developed for better inference accuracy [15]. Chen et al. reported a distance-racing current-mode sense amplifier to improve accuracy [5]. Ma. et al. proposed using extra devices to alleviate the uncertainty statistically [17]. However, these techniques require extra hardware to enhance the accuracy of the inference engine.

Software-based solutions are also proposed for specific ANVM-based CIM architectures, without adding complexity to the hardware design. Long et al. proposed dynamic fixed-point data representation and device variation aware training method to improve the network accuracy in a semi-analog architecture, where each analog memory device has a finite 2-state precision [16]. However, the studied CIM architecture does not fully utilize the high accuracy of some emerging ANVM devices such as CTT and PCM. In addition, the DNNs studied are not full-fledged; some standard operations of the state-of-the-art DNNs, including batch normalization [10] and residual filters [9], are not considered.

To pave the path for large-scale ANVM-based CIM inference engine for high-precision ANVM devices, where each analog device can be used for a high-precision synaptic weight, *we propose a new simulation platform* based on the PyTorch deep learning framework to evaluate the effect of analog device error from characterized hardware data. Based on the platform, we find that DNNs trained in digital engines (e.g., GPUs and CPUs) are not resilient to the ANVM device uncertainty without significant network redundancy in terms of extra synaptic weights.

To enhance the resiliency of DNNs, we propose the Hessian-Aware Stochastic Gradient Descent (HA-SGD) algorithm for neural network training with the analog floating-point precision, which can be used to enhance network resiliency without extra hardware cost. We also provide visualization for the networks trained by HA-SGD to cross-validate the resiliency improvement. In addition, since the analog-to-digital converters can consume a significant amount of the power and area in such architecture, we also evaluate the requirement of digitization effort in the mixed-signal CIM architecture.

# 3 Background

## 3.1 Analog Compute-in-Memory (CIM) Architecture for Neural Networks

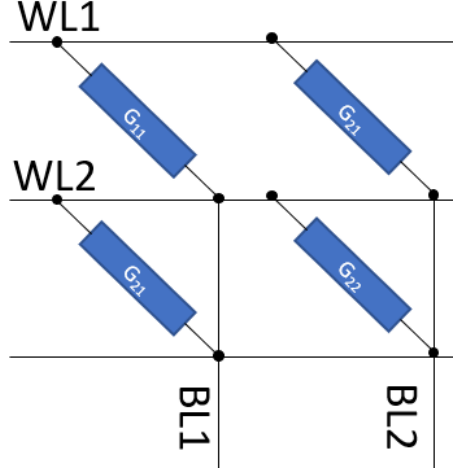

**Fig. 1 Crossbar array structure of a 2x2 array as an analog CIM architecture for vector-matrix multiplication**

Analog devices can be arranged into an array to build an analog compute-in-memory (CIM) vector-matrix multiplication (VMM) engine for applications such as neural networks [14]. A basic VMM is defined as $y = xW$ where $x$ is an input vector of 1 * M, $W$ is a weight matrix of M * N. A schematic of the VMM engine based on 2-terminal ANVM device is shown in Fig. 1. In this architecture, the analog devices based on the Charge Trapped Transistors (CTT) [7] are arranged in an M * N matrix with M rows and N columns. Each ANVM device in the array represents a synaptic weight in the weight matrix using its conductance $G_{mn}$ at a given WL, BL bias condition. Each word-line (WL) connects the input terminals of the devices in a row. Each bit-line (BL) connects the output terminals of the devices in a column. The input values can be encoded as pulse-width modulated (PWM) signals $V_m(t) = V_{on}[u(t) - u(t - x_m \Delta_t)]$ at the WLs (e.g., the $m$th row), where $u(t)$ is the step function, $V_{on}$ is the "on" voltage ("off" voltage is assumed to be 0V), $x_m$ is the $m$th entry of the input vector and $\Delta_t$ is the unit pulse width when $x_m = 1$. For the time when the input is "high", the $m$th device in the $n$th column will draw the amount of charge $Q_{mn} = I_{inf,mn} x_m \Delta_t$ from WL to BL where $I_{inf}$ is the inference current of the device, which is the read current when the device in "on". $Q_{mn}$ is linearly proportional to multiplication result between the input value $x_m$ and the stored matrix value $I_{inf,mn}$. By charge conservation, the total charge moved to the BL of each column is the dot-product between the input and all the devices at that column:

$$Q_n = \left(\sum_m I_{inf,mn} x_m\right) \Delta_t$$

In addition, this VMM engine can be directly used for VMM with a bias term by adding an extra input and an extra row of devices in the array:

$$y = xw^T + b = [x, 1] * \begin{bmatrix} w_1 & b_1 \\ w_j & b_j \end{bmatrix}^T$$

where an extra input of "1" is used in the input with the extra weights programmed as the bias terms $b_1, \dots, b_j$. The dot-product output stored as charge can be further processed for activation functions if required. The charge can be discharged by a constant current source so that it becomes another PWM signal and can be used as the input to another array (i.e., the next layer of the network) as $t_n = Activation(\frac{Q_n}{I_{discharge}})$. The output PWM signal can be digitized by a counter as an option.

Although the device conductance is always a positive value, synaptic weights of many neural networks require bipolar range. This can be addressed by adding a reference device to each row of the array and take the difference between the inference current of the target device and reference $I_{inf} = I_{inf,device} - I_{inf,ref}$. Similarly, negative inputs can be realized by applying a 2-segment PWM input where

$$Q_n = Q_{n,PWM} - Q_{n,ref} = Q_n = \left( \sum_m I_{inf,mn} x_{m,PWM} - \sum_m I_{inf,mn} x_{m,ref} \right) \Delta_t$$

This CIM architecture assumes sufficient device programming accuracy so that each synaptic weight is represented by one ANVM device. However, it is also possible to use multiple ANVM devices to represent a synaptic weight to decrease the relative error. In one example, the ANVM device is used for two states, and multiple devices are arranged together as a multi-bit value. Multiplication using this value as an operand would be computed for each device and accumulated as a weighted sum $y = \sum_{k=1}^{K} 2^{k-1} y_k$ where $y_k$ is the partial summation due to the device of most-significant bit (MSB) of this k-device value, and $y_1$ is that of the least-significant bit (LSB). The summations can be done conveniently once all partial summations can be digitized and the $2^{k-1}$ scaling can be done by left-shifting the digitized value. However, digitizing of the analog values in the analog CIM engine using analog-to-digital converters (ADCs) can be costly in terms of chip area, latency, and power efficiency. Therefore, it is also important to study the necessity of high-resolution ADCs in the analog CIM engine.

## 3.2 Analog Device Uncertainty

In the proposed analog CIM architecture, since the computing is based on physical quantity (i.e., charge) and physical laws (i.e., charge integration and charge conservation) instead of symbolic representation (i.e., numbers) and arithmetic laws (addition and multiplication), the precision of the analog computing is potentially infinite (or only limited by the elementary charge) by nature. However, the precision of the analog system is limited by the non-ideality of the hardware manifested by the intrinsic error of the analog device and circuit. In this paper, we focus on the uncertainty of the ANVM devices, which is intrinsic to the ANVM technology and can be detrimental to the accuracy of any neural network mapped on to the ANVM-based hardware platform.

For the emerging ANVM devices, their uncertainty in terms of the programmed infinite-precision analog state (e.g., current under some fixed bias) is due to several reasons:
(a) Programming process. Since the programming mechanism of many emerging ANVM devices is stochastic, they require a closed-loop verification process to determine whether the programming is finished. The criterion for the termination of the programming process depends on the user's ability to fine-tune the device and the precision of the measurement hardware. The non-ideal termination of the programming will affect the accuracy of programming.
(b) Device and cycle variation. For devices that use an open-loop process for programming, the variation among the devices and among the different cycles of the same device will contribute to the error of the programming.
(c) Imperfect data retention. The analog state of the device will keep changing over time due to various reasons related to its programming mechanism. It is important to predict the time of failure for the ANVM-based systems and suggest solutions such as data refreshing. However, the refreshing process itself will again suffer from the programming errors.

Many emerging ANVM technologies are characterized with optimized programming strategies. The device uncertainty can be modeled as a Gaussian distribution as shown in **Fig. 2**. [2, 7, 8, 25]. This error model of the device needs to be converted to the error model of synaptic weights. The synaptic weight matrices $W$ can thus be written as $W = W_0 + \Delta W$ where $W_0$ is the ideal synaptic weights obtained from the training engine and $\Delta W$ is the error model of the weights. Since weight-current conversion is normally done by linear mapping, $\Delta W$ can also be represented as a Gaussian random variable $\Delta W \sim N(\mu, \sigma^2)$ to reflect the device error characterized from the hardware. As this error is a stochastic property of the device itself, its variation cannot be corrected by other circuits and will inevitably propagate to all calculations. Therefore, it is crucial to evaluate whether a network can still be useful with the ANVM device uncertainty for the evaluation the feasibility of an ANVM technology for the mixed-signal CIM inference engine.

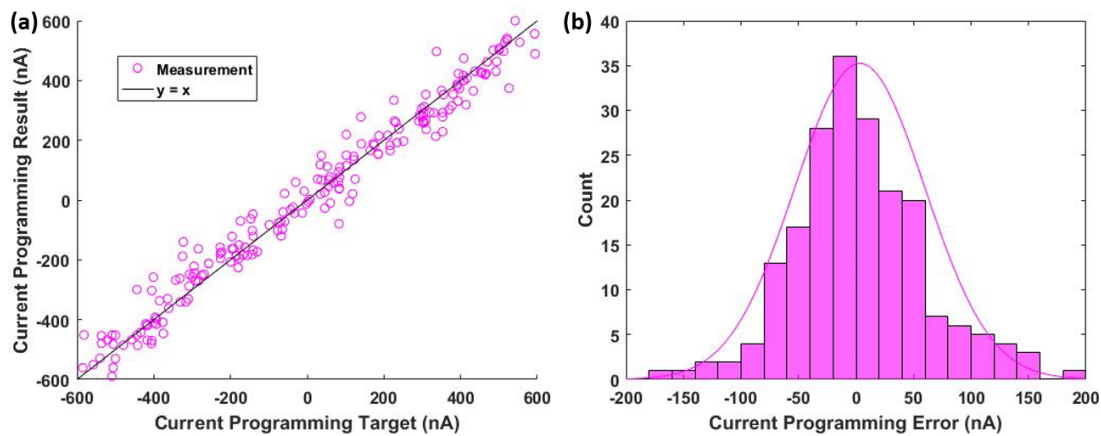

**Fig. 2 Typical ANVM device programming error characteristics (adapted from [7]) in scatter plot (a) and histogram (b).**

## 4 Hardware-Aware Evaluation and Training of Neural Networks

### 4.1 Analog Floating-Point Representation

To map the desired network synaptic weights efficiently to the device analog states, the analog floating-point precision is proposed based on the CIM hardware architecture. Since the computation and communication between filters are based on charge, the output values of a computation can be written as $t_{out} = Q_{out}/I_{discharge}$, where $I_{discharge}$ is the discharging current and $t_{out}$ is the converted pulse-width modulated signal to the next layer. This form can be regarded as an analog floating-point precision where $Q_{out}$ is the base value and $I_{discharge}$ is a multiplier that can be different for different filters or layers.

The analog floating-point representation can be accompanied by the device uncertainty model to establish a simulation framework for the analog CIM engine. First, the device noise is defined as the ratio between the standard deviation of the $I_{inf}$ error (as in a Gaussian model) and the entire range of device $I_{inf}$, which is around 4% - 7% for the ANVM cell based on CTTs [22], and can be similar or more for the other analog devices [2, 8, 25]. Then the device noise is used to map device error from the physical domain onto the synaptic weights in the numerical domain.

First, the memory windows of the cells used in the CIM engine need to be characterized. For example, the implementation of the bipolar weights requires a reference to define the differential current, which can be another ANVM device programmed to the middle point of its memory window, and the memory window of the twin-device cell is symmetric about zero. Then the maximum and minimum synaptic weight of each filter (i.e., an array) of the neural network is calculated. Since the maximum is always positive and the minimum is always negative, the one with larger magnitude (i.e., $w_{absmax} = \max(w_{max}, |w_{min}|)$) is used, and this value is mapped to the maximum of the window. This makes sure the memory window is properly used. As a result, the mapping coefficient between the device conductance and the weight is $\beta = \frac{G_{max}}{w_{absmax}} > 0$, which is constant in one filter of the neural network, but can be different in different filters due to different $w_{absmax}$ values. Suppose filter A and B are different and $w_{absmax,A} \neq w_{absmax,B}$ then $\beta_A = \frac{G_{max}}{w_{absmax,A}}$ and $\beta_B = \frac{G_{max}}{w_{absmax,B}}$, $\beta_A \neq \beta_B$ since the $G_{max}$ can be defined as the same across the system for the ease of device programming.

The different mapping coefficient $\beta$ does not affect the result of the computation in the numerical domain when the activation function (if exists) applied to the filter's output is linear in both positive and negative domains such as rectifying linear unit (ReLU) and leaky ReLU. In these cases, the mismatch of $\beta$ can be compensated in hardware by adjusting the magnitude of the discharging current during the output PWM signal generation. Suppose a synaptic weight $w_A$ in filter A has the same numerical value with a synaptic weight in filter B, i.e., $w_B = w_A$. Then for the identical input of $x$, the multiplication output $y$ must be equal: $y_A = xw_A = xw_B = y_B$. The input is denoted as $t_x$ since it is the time duration of the PWM signal, then the charge accumulated by this computation is $Q_A = V_{on}G_A t_x$, $Q_B = V_{on}G_B t_x$, where $V_{on}$ is the bias to turn the device "on". The mapping is $G_A = \beta_A * w_A$ and $G_B = \beta_B * w_B$. When $\beta_A \neq \beta_B$, $G_A \neq G_B$ but $G_A$ and $G_B$ will be at the same polarity. Then $t_{y_A} = f(\frac{Q_A}{I_{discharge,A}})$ and $t_{y_B} = f(\frac{Q_B}{I_{discharge,B}})$, where $f$ is the activation function. The computation in the numerical domain, $y_A = w_A * x = w_B * x = y_B$, requires that $t_{y_A} = t_{y_B}$ in the physical domain. Therefore $t_{y_A} = f(\frac{Q_A}{I_{discharge,A}}) = t_{y_B} = f(\frac{Q_B}{I_{discharge,B}})$, which will always hold if $\frac{\beta_A}{I_{discharge,A}} = \frac{\beta_B}{I_{discharge,B}}$.

This indicates that the mapping coefficient is controlled by the discharge current, which is the scaling factor of the analog floating-point representation. As a result, the Gaussian error model of the analog device conductance (i.e., $I_{inf}/V_{on}$) is mapped to the Gaussian error model of the synaptic weights by $\Delta w = \Delta G/\beta$ for simulation. In each network filter, $\Delta W$ is the error of the weight matrix in which all entries are sampled from the independent and identically distributed random variable $\Delta w$. Since $\Delta G$ is a known statistics, $\beta$ should be set as large as possible, and can be different for different layers to minimize the disturbance from $\Delta w$ at each layer.

## 4.2 Simulation Framework

The proposed simulation framework is based on the PyTorch deep learning framework and implements the analog floating-point representation of the synaptic weights. It is compatible with common training techniques (e.g., $\ell 1/\ell 2$ regularization, dropout, etc.). After a network is trained, the simulator is used to evaluate the performance of the ANVM-based DNNs. During the forward-propagation, perturbations for all weights from all layers are sampled from the specified random distributions before each test run. The same set of sampled weights is used for the entire test set to generate the accuracy scores. Since the weights are now stochastic, multiple testing runs are performed to obtain the statistics of the accuracy scores for a given ANVM based DNN. The following describes the implementation of the DNN building blocks and operations that is compatible with the proposed CIM engine and non-zero uncertainty of the weights.

### 4.2.1 Convolutional layers and Fully connected layers

Independent random noise of a given distribution is sampled and used to perturb the weights of one layer based on the analog floating-point representation. We define the device shift and device noise parameters $\mu_{DS}, \sigma_{DN}$ as the ratio between the mean ($\mu_{device}$) of the device error and the dynamic range of the device $\mu_{DS} = \frac{\mu_{device}}{Range_{device}}$, and as the ratio between the standard deviation ($\sigma_{device}$) of the device error and the dynamic range of the device $\sigma_{DN} = \frac{\sigma_{device}}{Range_{device}}$. Both parameters can be directly used in the numerical domain $\mu_{DS} = \frac{\mu_{weight}}{Range_{weight}}, \sigma_{DN} = \frac{\sigma_{weight}}{Range_{weight}}$.

When applicable, the bias terms of the layers, are combined with the weights and in the VMM engine for the ease of hardware implementation as previously proposed. However, the range of the bias parameters is often found to be higher than that of the weight parameters. Simply combining the bias and weight together will make the parameter range higher than the range of weights, leading to asymmetrically high noise levels for the weights. This could have a huge influence on the inference accuracy.

Therefore, we scale the bias terms by the extra dummy input. Instead of using "1" as the extra input for the bias term, the extra input can be other value to correspondingly scale bias parameters to match the range of the programmed biases with the range of weights:

$$y = xw^T + b = [x, s] * \begin{bmatrix} w_1 & b_1/s \\ w_j & b_j/s \end{bmatrix}^T$$

where $s \geq 1$ is a proper scaling factor. This ensures that both weights and biases use the full dynamic range of the device (**Fig. 3**) to minimize the effect of the device noise and device shift on the weights. **Fig. 4** shows the performance of the neural networks with and without matching the range of weights and biases.

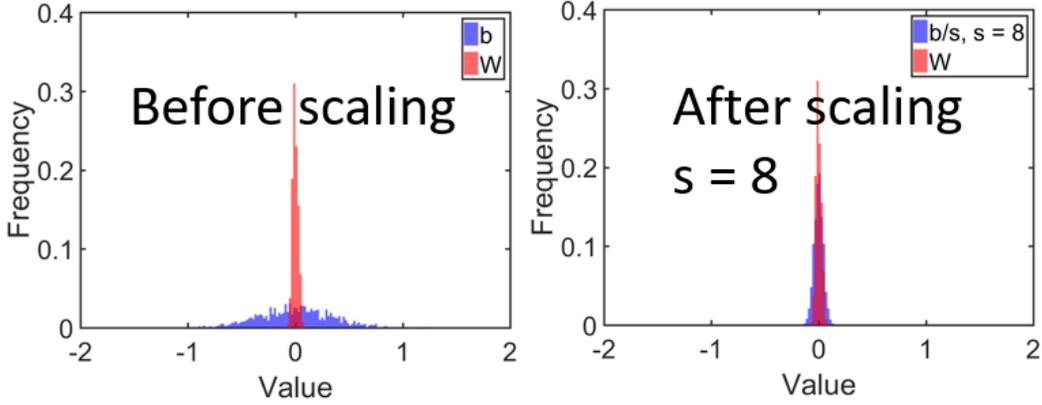

**Fig. 3 Normalized histograms showing the distribution of the weights *w* and bias *b* of a filter before and after**

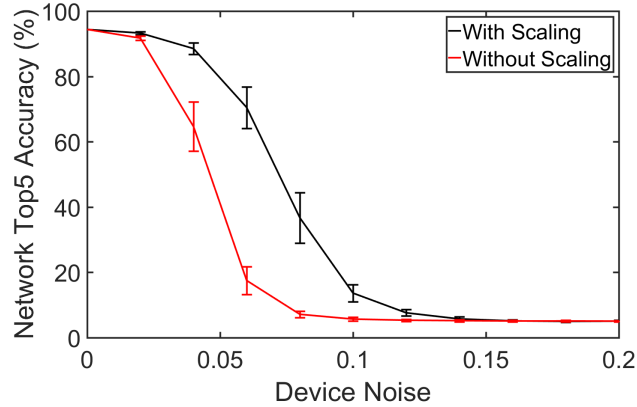

**Fig. 4 Network (Wide-ResNet-28) top5 accuracy on CIFAR-100 with and without the scaling**

### 4.2.2 Batch Normalization layers

Batch normalization is essentially another linear operation, which normalized the input to each channel individually with learned parameters (mean $\mu_c$, variance: $\sigma_c^2$, learnable scaling factor $\gamma_c$ and learnable bias $\beta_c$, for each channel c):

$$y_{c,i} = w_{eff,c} \, x_{c,i} + b_{eff,c}$$

where $w_{eff,c} = \frac{\gamma_c}{\sqrt{\sigma_c^2 + \epsilon}}$, $b_{eff,c} = \beta_c - \frac{\gamma_c \mu_c}{\sqrt{\sigma_c^2 + \epsilon}}$. Therefore, it can be implemented through convolutional layers with unit-size, unit stride convolutional kernel, which has the weight $w_{eff,c}$ and bias $b_{eff,c}$.

### 4.2.3 Shortcut layers and Residual blocks

Shortcut layers were introduced by ResNet [9] to address the gradient vanishing problem, and have become an indispensable component in deep neural networks. In our noise-considering implementation of shortcut connections we assumed that each positive entry of the identity matrix suffers from a Gaussian noise U with zero mean and variance δ, where δ is the device noise, and $I_{shortcut} = U + I$. This is to reflect the noise from possible circuit implementation of the shortcut (e.g., current mirror).

Residual blocks used in ResNet are implemented based on the convolutional layers, batch normalization layers and shortcut layers. The parameters for the behavior of analog devices (e.g., mean and variance in the case of Gaussian noise) can be individually specified for each layer.

### 4.2.4 Digitization layers

To evaluate the effect of digitization in the mixed-signal CIM engine (lower resolution than the 32-bit floating-point precision of the simulator), the quantization layer is designed and can be inserted after the activation functions, to represent the optional analog-to-digital converter (ADC) in the mixed-signal CIM engine.

## 4.3 Hessian-Aware Stochastic Gradient Descent (HA-SGD)

To improve the analog resiliency of the DNNs at the software level, which is in parallel with the hardware and device improvements, we introduce a Hessian-Aware Stochastic Gradient Descent (HA-SGD) algorithm to ensure that at convergence the local minimum will not have high-norm Hessian. At any given weight $W_0$, the HA-SGD samples in the weight space around $W_0$ by random perturbations, and uses the average of the gradients at the sample points as an estimated gradient of the smoothened cost function ($\tilde{J}(W) = E_{W'}[J(W')]$) at $W_0$. The variance of the random perturbations of the network weights is referred to as the level or intensity of the training noise in HA-SGD. The training algorithm samples the local neighborhood (around $W_0$) of the loss landscape and seeks local minima in the smoothened cost function, which correspond to wide valleys (with low-norm Hessian) in the original landscape. As a result, in the wide valleys, the neural network perturbed by finite device noise would maintain performance that is comparable to the case having no noise, with small variance and bias.

We present and provide motivation for our HA-SGD algorithm, and provide arguments for why it is able to avoid local minima with a high-norm Hessian (i.e., a Hessian with a large trace or, equivalently, large eigenvalues). First, we make intuitive arguments: It is useful to observe that the expectation of the cost function (under random perturbations of the weights) can be seen as the convolution between the original cost function and a kernel defined by the distribution of the random noise Δw:

$$E_{p(\Delta w)}[J(w + \Delta w; x)] = \int J(w + \Delta w; x) \, p(\Delta w) \, d\Delta w \coloneqq \tilde{J}(w; x)$$

Thus, the expected cost function $\tilde{J}(w; x)$ is a smoothened version of the original cost function, and could be treated as our new cost function that can be minimized to train the DNNs to be deployed in ANVM-based CIM engines.

Optimizing this smoothened cost function is compatible with the pursuit of a set of weights, the random perturbations around which lead to low expected-value of loss with low variance. Narrow local minima, which have high-norm Hessians, will be shallow local minima or even vanish in this smoothened new cost function, and thus be avoided. Any local minimum of $\tilde{J}(w; x)$ is Hessian-Aware and steep valleys are naturally penalized. Thus, the low-variance objective and low-expectation objective can be jointly pursued by searching for a good local minimum of $\tilde{J}(w; x)$. However, because of the extremely high dimensionality of **w** in DNNs, the estimation of the expectation can be of high variance, and it becomes impractical to optimize the exact $\tilde{J}(w; x)$. Theoretical guarantees remain unclear whether a stochastic gradient algorithm for optimizing the Hessian-Aware cost function $\tilde{J}(w; x)$, will, in fact, avoid steep valleys and penalize for high-trace Hessians.

Below we motivate our HA-SGD algorithm by minimizing $\tilde{J}(w; x)$, and explain why it would penalize any minimum with large-norm Hessians. Using this new cost function, the goal is to solve for

$$w^* = \arg\min_w \tilde{J}(w; x)$$

To optimize $\tilde{J}(w; x)$, we calculate its gradient with respect to w:

$$\nabla_w \tilde{J}(w; x) = \nabla_w E_{p(\Delta w)}[J(w + \Delta w; x)]$$

Under mild assumptions for applying the Leibniz integral rule, the expectation operator and derivative operator can be swapped, giving

$$\nabla_w \tilde{J}(w; x) = E_{p(\Delta w)}[\nabla_w J(w + \Delta w; x)]$$

The gradient of an expectation is converted to the expectation of a gradient, and can be estimated by the sample mean of L samples:

$$\nabla_w \tilde{J}(w; x) \approx \frac{1}{L} \sum_{\ell=1}^{L} \nabla_w J(w + \Delta w^{(\ell)}; x)$$

The weights are then updated using this gradient,

$$w \leftarrow w - \eta \frac{1}{L} \sum_{\ell=1}^{L} \nabla_w J(w + \Delta w^{(\ell)}; x)$$

where η is the stepsize.

For stochastic gradient descent,

$$\nabla_w \tilde{J}(w; x^{(i)}) \approx \frac{1}{L} \sum_{\ell=1}^{L} \nabla_w J(w + \Delta w^{(\ell)}; x^{(i)}),$$

$$w \leftarrow w - \eta \frac{1}{L} \sum_{\ell=1}^{L} \nabla_w J(w + \Delta w^{(\ell)}; x^{(i)}).$$

We now provide a first-order analysis showing that the algorithm will not converge to a **w** with a high-norm Hessian. To simplify the analysis, we assume the perturbation $\Delta w$ is zero-centered. The non-zero-centered case will only differ by a global shift. Expanding $\nabla_w J(w + \Delta w; x)$ to the first-order gives

$$\nabla_w J(w + \Delta w; x) = \nabla_w J(w; x) + H(w)\Delta w + O(\Delta w^2).$$

Then

$$\nabla_w \tilde{J}(w; x) \approx \nabla_w J(w; x) + \frac{1}{L} H(w) \sum_{\ell=1}^{L} \Delta w^{(\ell)}.$$

In the first-order approximation, this adds a Hessian-related perturbation to the ordinary gradient at w. The variance of such random perturbation is $\frac{1}{L}\text{tr}(\text{Var}[H(w)\Delta w]) = \text{tr}(H(w)\Sigma H(w)^T)$. Thus, the optimizer will never "settle down" until it finds a local minimum where the gradient vanishes, *and* the Hessian-related perturbation is always low.

Intuitively, in the first-order, the optimizer probes the surrounding landscape with several samples, infers a local parabolic approximation of the landscape and uses that information to adjust the gradient descent step to find flatter local minima.

For realistic implementations of analog DNNs, because the range of the synaptic weights is mapped to the dynamic range of the CTT devices, and the uncertainties are proportional to the dynamic range, the standard deviation of $\Delta w$ is proportional to the range of the weights programmed. That is,

$$\Delta w = \alpha(w_{max} - w_{min})(\epsilon + \mu)$$

where $\epsilon \sim p(\epsilon)$ is the basic form of the zero-centered random noise that is independent of $w$, and $\mu$ is the constant shift part of the random noise. In this case,

$$\nabla_w J(w + \Delta w; x) = \nabla_w J(w + \alpha(\epsilon + \mu)(w_{max} - w_{min}); x)$$

$$= \left(\frac{\partial(w + \alpha(\epsilon + \mu)(w_{max} - w_{min}))}{\partial w}\right)^T \nabla_\omega J(\omega; x)\bigg|_{\omega = w + \alpha(\epsilon + \mu)(w_{max} - w_{min})}$$

and

$$\left(\frac{\partial\bigl(w+\alpha(\epsilon+\mu)(w_{max}-w_{min})\bigr)}{\partial w}\right)_{ij} = \frac{\partial\bigl(w+\alpha(\epsilon+\mu)(w_{max}-w_{min})\bigr)_i}{\partial w_j} = \begin{cases} \delta_{ij}+\alpha(\epsilon+\mu), & j=i_{max} \\ \delta_{ij}-\alpha(\epsilon+\mu), & j=i_{min} \\ \delta_{ij}, & otherwise \end{cases}$$

Where $\delta_{ij}=\begin{cases}1, & i=j \\ 0, & i\neq j\end{cases}$. Thus, even in reality the random perturbations are not completely independent of the weights, the gradient form remains simple and very close to the gradient form when assuming such independence. That is,

$$\nabla_w J(w+\Delta w; x) = \nabla_w J(w+\alpha(\epsilon+\mu)(w_{max}-w_{min}); x) \approx \nabla_\omega J(\omega; x)|_{\omega=w+\alpha(\epsilon+\mu)(w_{max}-w_{min})}$$

Thus the analysis of HA-SGD above remains practical.

## 5 Results

### 5.1 Analog Resiliency of Neural Network

The simulation framework proposed in the last section is used to simulate DNNs based on Wide ResNet models [23]. It includes many of the advanced operations of state-of-the-art neural networks (e.g., batch normalization, residual blocks), which are affected by the device shift and device noise. The DNN model is trained on commercial GPUs using 32-bit floating-point precision and the CIFAR-100 training set. During testing of the trained network, device models are included to emulate the behavior of analog devices. Each pre-trained digital network is instantiated 50 times by independent sampling from the device uncertainty statistics (i.e., $w = w_0 + \Delta w$). Then the system is evaluated by all testing patterns on all instantiated networks to obtain the statistics for network accuracy.

Two different structures of Wide ResNet with depth level 16 (17.1 million weights) and 28 (36.5 million weights) are trained and tested on the CIFAR-100 dataset (same for all network simulation results shown unless otherwise specified). We first examine the case where $\mu_{DS}=0, \sigma_{DN}>0$. In **Fig. 5**, all networks presented show significant degradation of the network accuracy as the $\sigma_{DN}$ increases, while the larger network is more resilient. When $\sigma_{DN}=6\%$, the top5 accuracy is degraded from 94.28% to 44.96% for the 16-layer network and from 94.39% to 70.99% for the 28-layer network. Both networks start to fail completely at $\sigma_{DN}=14\%$. For reference, $\sigma_{DN}$ of some reported state-of-the-art ANVM can be from 5% to 20% [8, 22, 25]. The deeper network (also with more weights) is less sensitive to the increase of $\sigma_{DN}$, and therefore is more resilient to $\sigma_{DN}$. The resiliency of the network also depends on the application. **Fig. 6** shows the network accuracy of the 16-layer Wide ResNet on the CIFAR-100 classification, compared with the MNIST classification, which is significantly easier than CIFAR-100. The resiliency of the neural network of a similar scale is better for easier applications.

Conventional methods during training to enhance network generalization, such as dropout and $\ell 2$ regularization, can be used to improve the resiliency of the network. **Fig. 7** shows the Wide ResNet-28 trained using different dropout factors ($D$) and different $\ell 2$ regularization factors ($L$). The result showed in **Fig. 5** and **Fig. 6** used D = 0.3 and L = 0.0005.

Then we look at the effect of the device shift. **Fig. 8** shows the ResNet-18 trained with optimized $D$ and $L$ and tested with $\mu_{DS} \neq 0, \sigma_{DN}=0$. While the device shift has a smaller magnitude than the device noise in state-of-the-art ANVM devices, it has a significant degrading effect on the classification accuracy.

Therefore, for the state-of-the-art ANVM devices, the neural networks trained with conventional enhancement methods might not be resilient to the device noise when the network was trained in a high-precision digital machine and optimized for it.

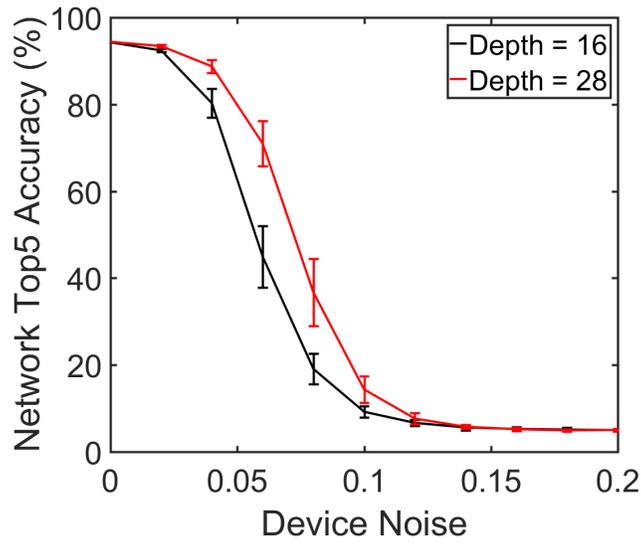

**Fig. 5 The degradation of network due to analog device noise: two Wide ResNet models of depth level 16 and 28 are trained and tested on the CIFAR-100 dataset.**

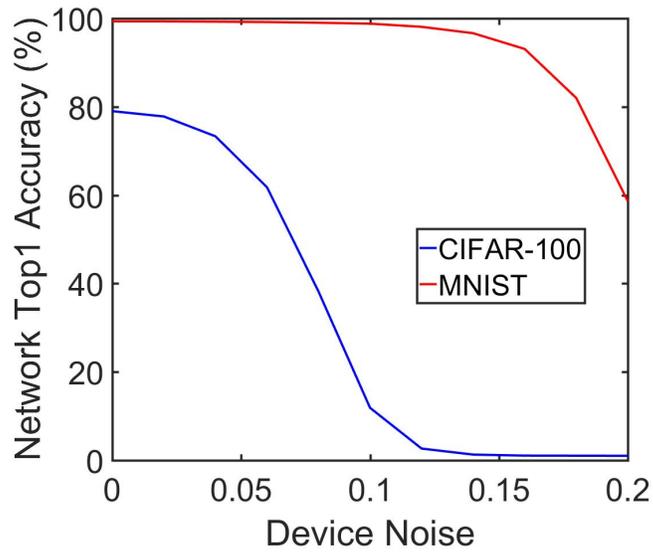

**Fig. 6 The degradation of network due to analog device noise for different applications: Wide ResNet model with depth level 16 for both CIFAR-100 and MNIST are trained and tested, showing that the network is more resilient to device noise when the application is simpler.**

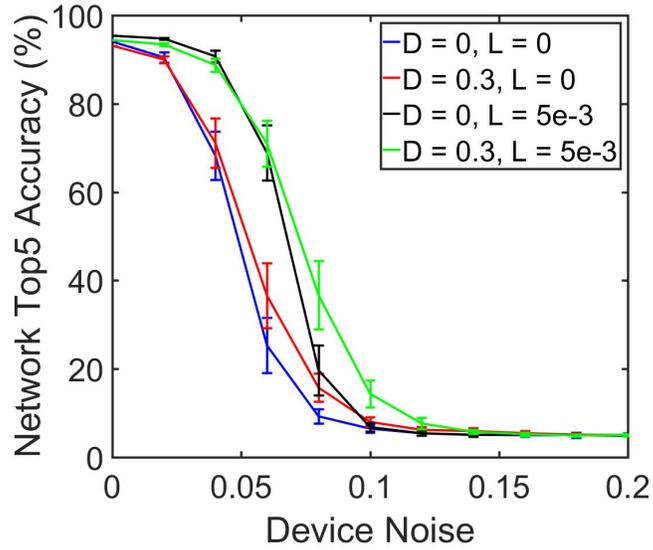

**Fig. 7 The degradation of network due to analog device noise using different dropout factors (*D*) and L2 regularization factors (*L*):** Wide ResNet model with depth level 28 for CIFAR-100 is trained and tested, showing that the network can be more resilient to device noise using these conventional generalization enhancement methods.

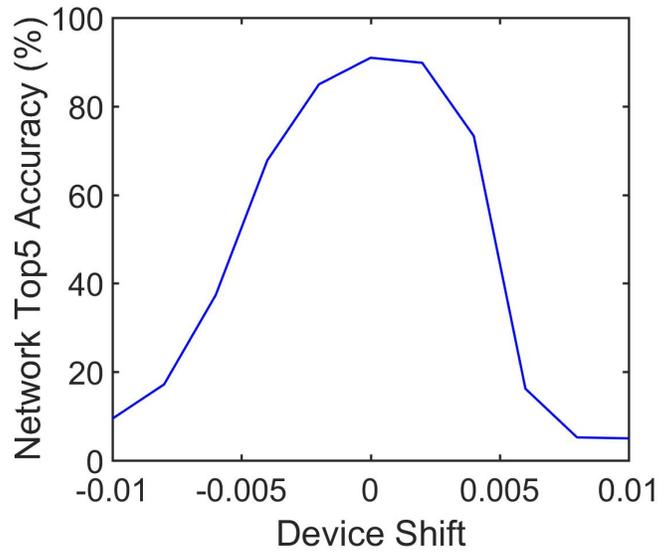

**Fig. 8 The degradation of network due to analog device shift ($\mu_{DS}$):** Wide ResNet-28 model for CIFAR-100 is trained using optimized *D* and *L* factors and tested

## 5.2 Improving Analog Resiliency by HA-SGD

The HA-SGD can improve the network resiliency significantly. **Fig. 9** shows the enhancement of resiliency when HA-SGD is used for training. At a 6% device noise, HA-SGD improves performance from 70.99% to 88.47%. HA-SGD has an even more significant effect when the device noise is higher. At 10% device noise, a network trained with injected noise of 10% achieves a top5 accuracy of 61.67%, which is more than four times the top5 accuracy of the networks trained by standard SGD (14.37%). The optimal level of the training noise depends on the device noise of the target device (e.g., CTT). **Fig. 10** shows the effect of increasing training noise on different device noise during inference. In general, a higher training noise performs better for higher device noise because the gradient estimation during training is more accurate. Notably, the network tested with zero device noise would not benefit from the HA-SGD algorithm, and therefore HA-SGD should only be applied when some finite device noise is expected in the analog inference engine. **Fig. 11** shows the network top1 accuracy for the Wide ResNet-28 tested for CIFAR-100. Three groups of networks are trained with different training noise (device noise injection in the forward propagation during training). The optimal accuracy can be achieved when the training noise is close to the expected device noise during testing. Also, in **Fig. 11** some degradation of the network at zero device noise is observed. This implies that the local minimum of the cost function found by HA-SGD is not necessarily deeper, despite being flatter.

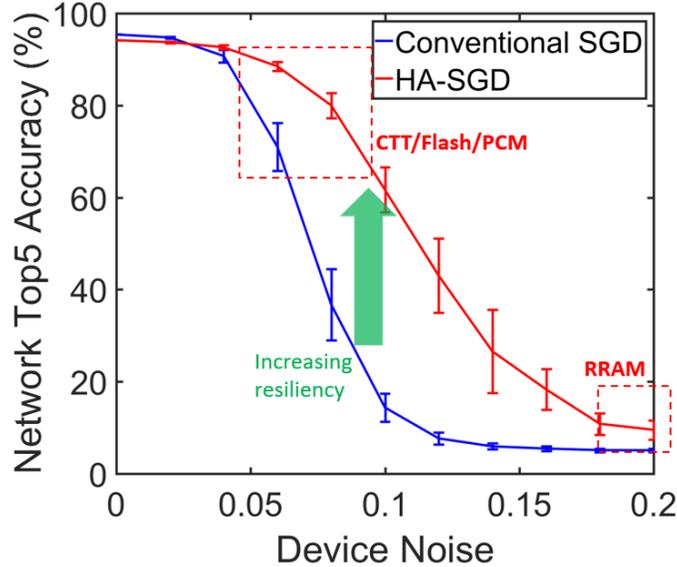

**Fig. 9** Improved network resiliency with HA-SGD: the performance of networks trained with the HA-SGD algorithm are compared with those trained by highly optimized conventional training algorithms, showing improved resiliency from the HA-SGD method. Parameters such as L2-regularization factor, dropout factor and the training noise level are all optimized for both cases. The device noise levels of some analog devices, such as CTT, Flash [8], phase change memory (PCM) [2], and resistive RAM (RRAM) [25] are indicated.

HA-SGD's preference of flatter minima and the associated degradation on zero-noise accuracy can be verified by evaluating the landscape of the loss function along the optimization trajectory during the course of training. In this experiment, we train a neural network (LeNet) on the MNIST dataset with different levels of training noise during HA-SGD, tracking the trajectories of their weights in the weight space. The full-batch loss and accuracy are evaluated for the trajectory points as well as isotropic Gaussian samples in the weight space along the trajectories to capture local landscape geometry. To visualize the trajectories in the high-dimensional weight space, we use principal component analysis (PCA) to project the trajectories to a 2D plane. **Fig. 12 (a)** shows the projected trajectories of four training runs with different levels of training noise, where a contour plot of the full-batch loss at the projected sample points is plotted as background. It shows that the HA-SGD with higher training noise levels would tend to converge towards a "wider" valley with a lower-norm Hessian. However, the loss function, when computed with $\sigma_{DN} = 0$, does not fall into the deeper valley when the training noise increases (**Fig. 12 (b)**), which corresponds to the accuracy degradation at $\sigma_{DN} = 0$ as shown in **Fig. 12 (c)**. In addition, the norm of the Hessian can be calculated using the largest eigenvalue of the covariance matrix of the batch gradient. This quantity is shown to be a good approximation of the Hessian's largest eigenvalue [19]. **Fig. 12 (d)** plots the evolution of this quantity along the trajectories, showing that with a higher perturbation level, HA-SGD goes to points with lower-norm Hessian in the weight space.

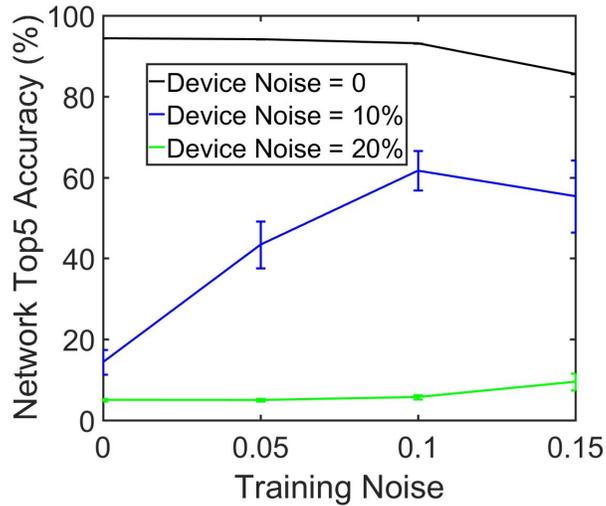

**Fig. 10 Effect of noise-level during training: Increasing the level of training noise in HA-SGD has different effect on the resiliency of the network, depending on the device noise of the target device.**

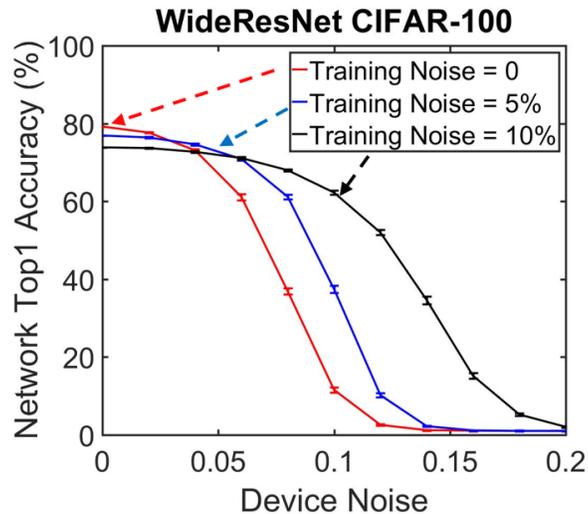

**Fig. 11 Network top1 accuracy for the WideResNet-28 tested for CIFAR-100. For a given device technology, the best network inference accuracy is achieved when the training noise is set to be close to the target device noise. As to be expected, the zero-noise inference accuracy might decrease as the training noise increases.**

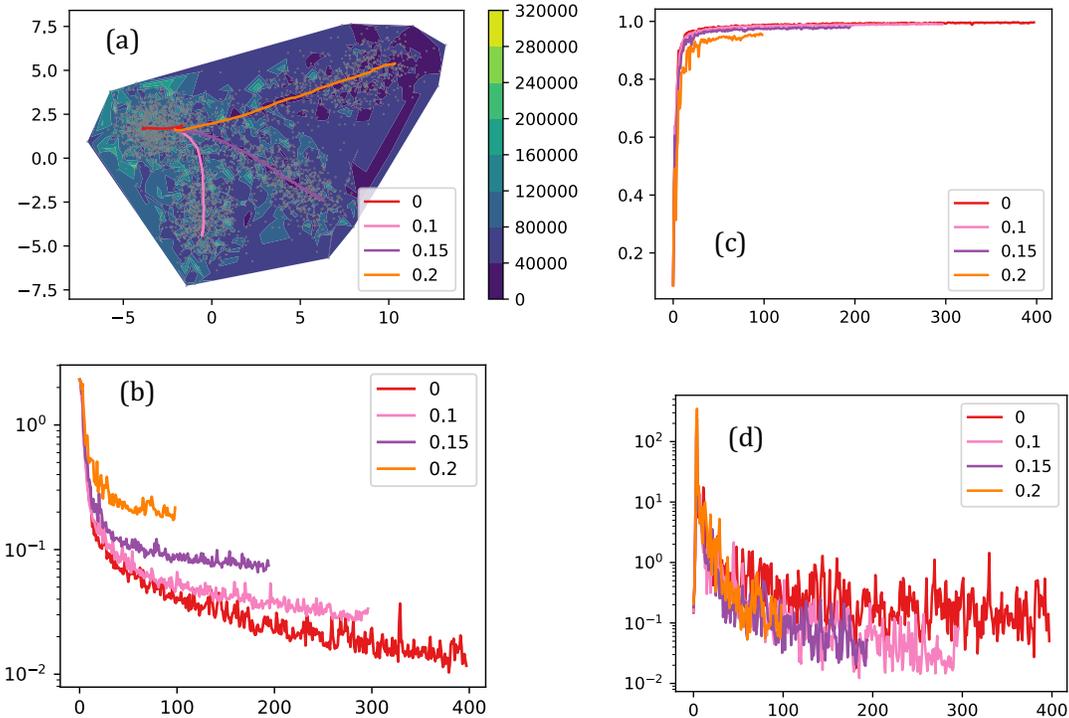

**Fig. 12 (a):** The PCA projection of trajectories of LeNet's parameters trained on MNIST with HA-SGD. For comparison, 4 networks are initialized for the same condition and trained by HA-SGD using different training noise:0, 0.1, 0.15, and 0.2 as denoted in the legend. The loss function values of isotropic Gaussian samples (the gray dots) near the trajectories are shown in the background contour plot. The loss function landscapes around the HASGD trajectories with higher training noise show less variation. The zero-training noise trajectory, on the other hand, is surrounded by wide fluctuations in loss function values, showing that the local minim is a steep one. b): The evolution of the loss during training. (c): The evolution of the accuracy of the networks during training (computed with $\sigma_{DN} = 0$). (d): The evolution of the largest eigenvalue of the gradient's covariance matrix, which approximates the network's Hessian's largest eigenvalue and quantifies the steepness of the local loss landscape. The zero-training noise case shows much higher values.

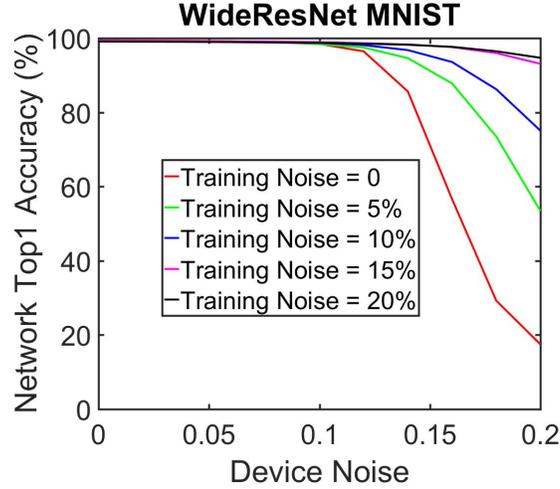

**Fig. 13 The zero-noise inference accuracy degradation can be less apparent and go unnoticed until the device noise is increased significantly, when the network is trained for a simpler problem, such as MNIST.**

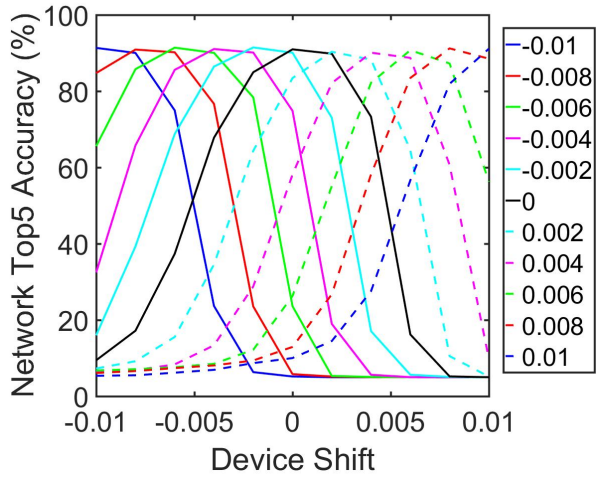

**Fig. 14 The degradation of network due to analog device shift: Wide ResNet model with depth level 28 for CIFAR-100 is trained and tested. Different networks are trained with different values device shift (showed in the legend). All networks are trained with 5% device noise and tested with $\sigma_{DN} = 5\%$. The mean shift of the device can therefore be compensated by using it as a prior knowledge during training.**

However, this degradation of the network due to HA-SGD can be hidden if a very deep network is trained for a simpler challenge. **Fig. 13** also shows the Wide ResNet-28 network trained for MNIST challenge using different training noise, whose accuracy at zero device noise during testing does not change.

The network resiliency to the mean shift of the device $\mu_{DS}$ can also be enhanced by adjusting the training strategy. Since the mean shift is not a random variable, the network can be trained with the addition of the mean shift to the weights. **Fig. 14** shows that the effect of mean shift for $\mu_{DN} < 1\%$ can be adapted by training if $\mu_{DS}$ can be characterized and used as prior knowledge during training. For devices whose $\mu_{DS}$ is a strong function of time, compensation in hardware such as refreshing the stored values as done in DRAM operations can also be helpful.

## 5.3 Quantization Resiliency of Neural Networks

In many mixed-signal CIM inference engine implementations, especially for convolutional neural networks, digitizing the output of layers is required to support the reuse of hardware resources. The digitization essentially injects a quantization error. In our simulation framework, we design a quantization layer for the quantization effect, which can be inserted at different stages of the neural network. *To best simulate the operation of the hardware, the quantization layer is placed after the activation operations to evaluate the effect of ADC resolution when used in the CIM architecture.* The device statistics reported for the CTTs is used for the simulation to capture both the effect of $\mu_{DS}$ and $\sigma_{DN}$ of real hardware [22].

**Fig. 15** shows the network accuracy as a function of quantization levels (for each quantization layer). It shows the saturation of the inference accuracy after 6-bit for the Wide ResNet-28 trained for CIFAR-100. The analog resiliency improvement due to HA-SGD is observed at almost all quantization levels and is more than 50% up to 8-bit quantization after each layer of the network.

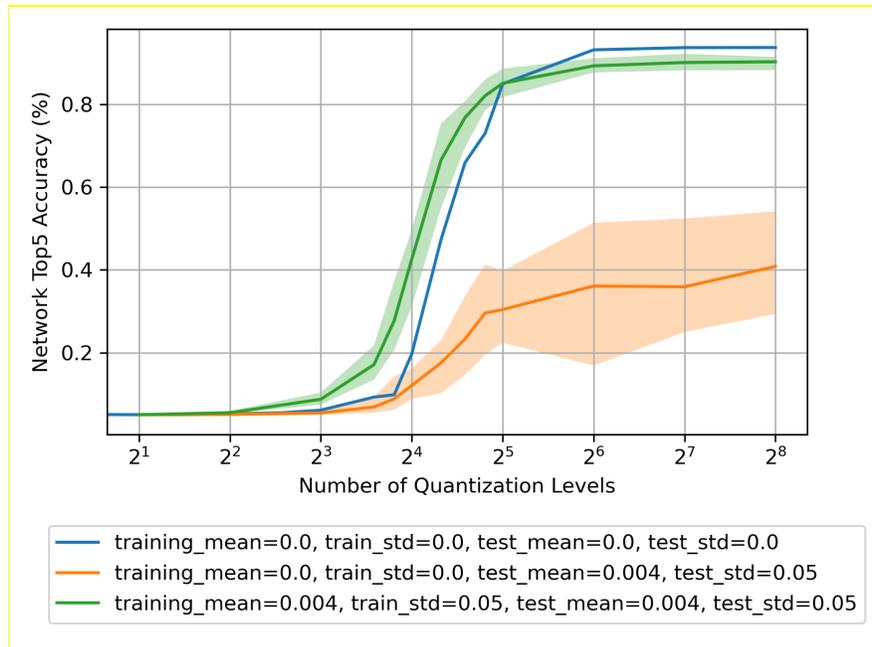

**Fig. 15** Effect of quantization levels for Wide ResNet-28 trained using different HA-SGD parameters on CIFAR-100 dataset

# 6 Concluding Remarks

This work has demonstrated a hardware-aware neural network deployment and training method to enhance the neural network's resiliency against errors in analog computing due to the ANVM characteristics based on CIM architecture. The analog floating-point representation of the weights is used to reduce the effect of device errors in the system. The Hessian-aware stochastic gradient descent (HA-SGD) is proposed for optimizing neural network training in analog CIM engines. The ANVM technology, once characterized for its error, can be used as prior knowledge and emulated during the forward propagation of the network training. Since one device corresponds to a weight in the proposed analog CIM engine, the error due to the device is a continuous random distribution in the weight domain. We have shown that the influence of this error can be ameliorated by the HA-SGD algorithm, where the expected gradient of the network (with the random errors in the weight) can be effectively approximately through sampling without explicit computation of the actual landscape of the loss function. The HA-SGD algorithm is shown to increase the accuracy of the neural network by up to 40%, and improve the network accuracy by more than 50% with quantization errors while requiring no extra cost in the inference hardware.